\documentclass[12pt,preprint]{aastex}
\shorttitle{PopIII Stars' Remnants} 
\shortauthors{Trenti, Santos \& Stiavelli}

\begin{document}


\title{Where can we really find the First Stars' Remnants today?}


%\author{The Three of Us}

\author{M. Trenti, M.~R. Santos, and M. Stiavelli} 
\affil{Space Telescope Science Institute, 3700 San Martin Drive Baltimore MD 21218 USA}
\email{trenti@stsci.edu; msantos@stsci.edu;  mstiavel@stsci.edu}


%-------------------------------------------------%
\begin{abstract}

A number of recent numerical investigations concluded that
the remnants of rare structures formed at very high redshift, such as
the very first stars and bright redshift $z\approx 6$ QSOs, are
preferentially located at the center of the most massive galaxy clusters
at redshift $z=0$. In this paper we readdress this question using a
combination of cosmological simulations of structure formation and
extended Press-Schechter formalism and we show that the typical
remnants of Population III stars are instead more likely to be found
in a group environment, that is in dark matter halos of mass $\lesssim
2 \cdot 10^{13}~h^{-1} \mathrm{M_{\sun}}$. Similarly, the descendants of the
brightest $z \approx 6$ QSOs are expected to be in medium-sized
clusters (mass of a few $10^{14}~h^{-1} \mathrm{M_{\sun}}$), rather than in the
most massive superclusters ($M>10^{15}~h^{-1} \mathrm{M_{\sun}}$) found within
the typical 1~Gpc$^3$ cosmic volume where a bright $z\approx 6$ QSO
lives.  The origin of past claims that the most massive clusters
preferentially host these remnants is rooted in the numerical method
used to initialize their numerical simulations: 
Only a small region of the cosmological volume of interest was
simulated with sufficient resolution to identify low-mass halos at
early times, and this region was chosen to host the most massive halo
in the cosmological volume at late times. The conclusion that the
earliest structures formed in the entire cosmological volume evolve
into the most massive halo at late times was thus arrived at by
construction. We demonstrate that, to the contrary, the first
structures to form in a cosmological region evolve into relatively
typical objects at later times. We propose alternative numerical
methods for simulating the earliest structures in cosmological
volumes.
 
\end{abstract}

\keywords{cosmology: theory - galaxies: high-redshift - early universe
- methods: N-body simulations}

%%%%%%%%%%%%%%%%%%%%%%%%%%%%%%%%%%%%
\section{Introduction}

Rare dark matter halos at high redshifts host interesting
astrophysical objects, especially before or at the end of the
reionization epoch. One example is given by the very first Population~III (PopIII)
stars formed in the universe at $z \gtrsim 40$, which started the metal
enrichment of the interstellar medium and the reionization process
\citep{abel02,san02,bromm04,naoz06}, and possibly also produced
intermediate mass black hole seeds that grow to become
super-massive black holes ($M_{BH}> 10^9 M_{\sun}$) within the first
billion year after the Big Bang \citep{vol03}. Another example are
bright $z \approx 6$ QSOs \citep{fan04}, considered to be hosted in the
most massive dark matter halos at that time \citep{MR05}. The
luminosity of such an object is powered through accretion onto a
supermassive black hole \citep[e.g. see][]{hop06}, which may be the
descendant of one of the first generation of PopIII stars formed in
the universe, at $z \gtrsim 40$ (\citealt{mad01}; see also \citealt{ts07a}).

Many numerical and theoretical investigations have aimed at
characterizing the properties of both PopIII stars
\citep[e.g. see][]{abel02,bromm04,ciar,ree05,gao05,osh07} and of high
redshift QSOs \citep[e.g. see][]{hop06,MR05,dimatteo05,li07}. However,
the rarity of these structures makes a fully self-consistent treatment
even of the formation of the underlying dark matter halos
typically outside the current capabilities of a
standard cosmological simulation, as the dynamic range that needs to
be resolved is too large. The main limitation at high mass resolution is the box size,
$l_{box}$: The enforcement of periodic boundary conditions
limits the power spectrum of density fluctuations to modes with
wavelengths shorter than $l_{box}$. This results in a severe bias on
the measured number density of rare, massive halos, especially
before the epoch of the reionization of the universe when the abundance
of galaxies derived from numerical simulations can be underestimated
by up to an order of magnitude \citep{bar04}. While the number density
of rare objects can be estimated in the context of
Press-Schechter-type modeling \citep{lacey93,mo_white96,she99,bar04},
studying the details of the formation histories of the first galaxies
and QSOs, possibly including hydrodynamics and radiative feedback
processes, requires high-resolution numerical simulation.

Thus, given a simulation box that is large enough to contain one or
more of the high-redshift halos of interest, the problem is to
identify a sub-region that contains one of them for high-resolution
simulation. In passing we note that this challenge is different from
that of developing a simulation code that is able to adaptively
increase the resolution when following the gas collapse that leads to
the formation of the first stars. This has been successfully
implemented with adaptive mesh refinement codes (for example ENZO -
\citealt{bry98}). \citet{gao05} proposed a method to identify rare
structures formed at very high redshift by recursively resimulating
successively smaller, nested, sub-regions of a simulation box at
progressively higher resolutions. Specifically, their method
resimulates the region of the box centered on the most massive dark
matter halo, reidentified at higher redshift and lower mass within the
sub-region at each resimulation step, until the desired resolution is
achieved in a small fraction of the entire simulation box. This
procedure is well-motivated by the fact that number density of massive
halos is increased in regions of large-scale over-density
\citep{bar04}, and thus structure formation in the sub-region on small
scales at early times is accelerated by sitting at the top of an
over-density, an over-density that is known to exist because it collapsed
into the massive halo identified in the previous resimulation
step. The \citet{gao05} method extends high-resolution resimulation
techniques used previously \citep[e.g. see ][]{nav94,whi00}, and has
been used recently by several authors to study the first stars and
QSOs \citep{gao05,ree05,gao07,li07}.

The recursive resimulation method for volume selection is very
effective at identifying a region with a very high-redshift halo when
compared to random selection of a region of equal volume within the
parent simulation box, but still it does not lead to the rarest
high-redshift halos in the volume. This is due to the stochastic
nature of the formation and growth of dark matter halos
\citep[e.g. see][]{PS,bond,she99}. The most massive dark matter halo
in a given volume at some early time does not in general evolve into
the most massive halo at a later time. The probability of the most
massive halo to evolve into the most massive halo, quantified
following \citet{lacey93}, decreases as time passes and the
characteristic mass scale increases. As we demonstrate, the
probability falls far below unity for the evolution of the first stars
and QSOs to the present time. The stochastic nature of growth
histories of dark matter halos is addressed and discussed in
\citet{gao05}, but unfortunately some of the subsequent studies using
their method do not take into account this stochasticity and assume a
very strong correlation between the location of the richest clusters
at $z=0$ and that of rare halos at very high redshift.

In this paper we clarify this issue by presenting some basic, but
often overlooked, results of Gaussian random fields in extended
Press-Schechter theory to quantify the locations at later times of the
first dark matter halos to form in a simulation box. We compare these
analytical and numerical results to those published using recursive
resimulation and highlight the benefits and limitations of that
method. In addition, we propose an alternative method, based on
analysis of the density field in the initial conditions, to select the
sub-region of a simulation box that contains some of the earliest
structure in the box without being biased toward the regions hosting
the largest halos at $z=0$. The paper is organized as follows. In
Sec.~\ref{sec:QSOs} we investigate where the descendants of $z=6$ QSOs
are today, while in Sec.~\ref{sec:PopIII} we extend the result to the
first Population III stars in the Universe. In Sec.~\ref{sec:Ic} we
discuss the implications of our results for the recursive mesh
refinement method and propose a viable
alternative. Sec.~\ref{sec:conc} summarizes and concludes.

\section{Bright $z\approx 6$ QSOs: where are they today?} \label{sec:QSOs}

As a first step toward characterizing the assembly history of rare,
massive, dark matter halos, we consider the link between $z \approx 6$
QSOs (considered to be hosted in the largest protoclusters at that
time, e.g. see \citealt{MR05}) and the largest clusters at $z=0$. For
this we use the publicly available merger trees constructed upon the
Millennium Run \citep{MR05,lemson}. This cosmological simulation has
more than $10^{10}$ dark matter particles within a volume of
$500^3~h^{-3}\mathrm{Mpc}^3$ and has been run using a concordance
$\Lambda CDM$ cosmology based on WMAP year 1 results
\citep{WMAP1}\footnote{Note that a change in the cosmological
parameters, and in particular on the value of $\sigma_8$, the root
mean squared mass fluctuation in a sphere of radius
$8~h^{-1}\mathrm{Mpc}$ extrapolated to $z=0$ using linear theory, does
influence the details of our results, such as the expected mass of a
QSO host halo, but not the essence of the relation between dark matter
halos at different redshifts.}.

\begin{figure} 
  \plotone{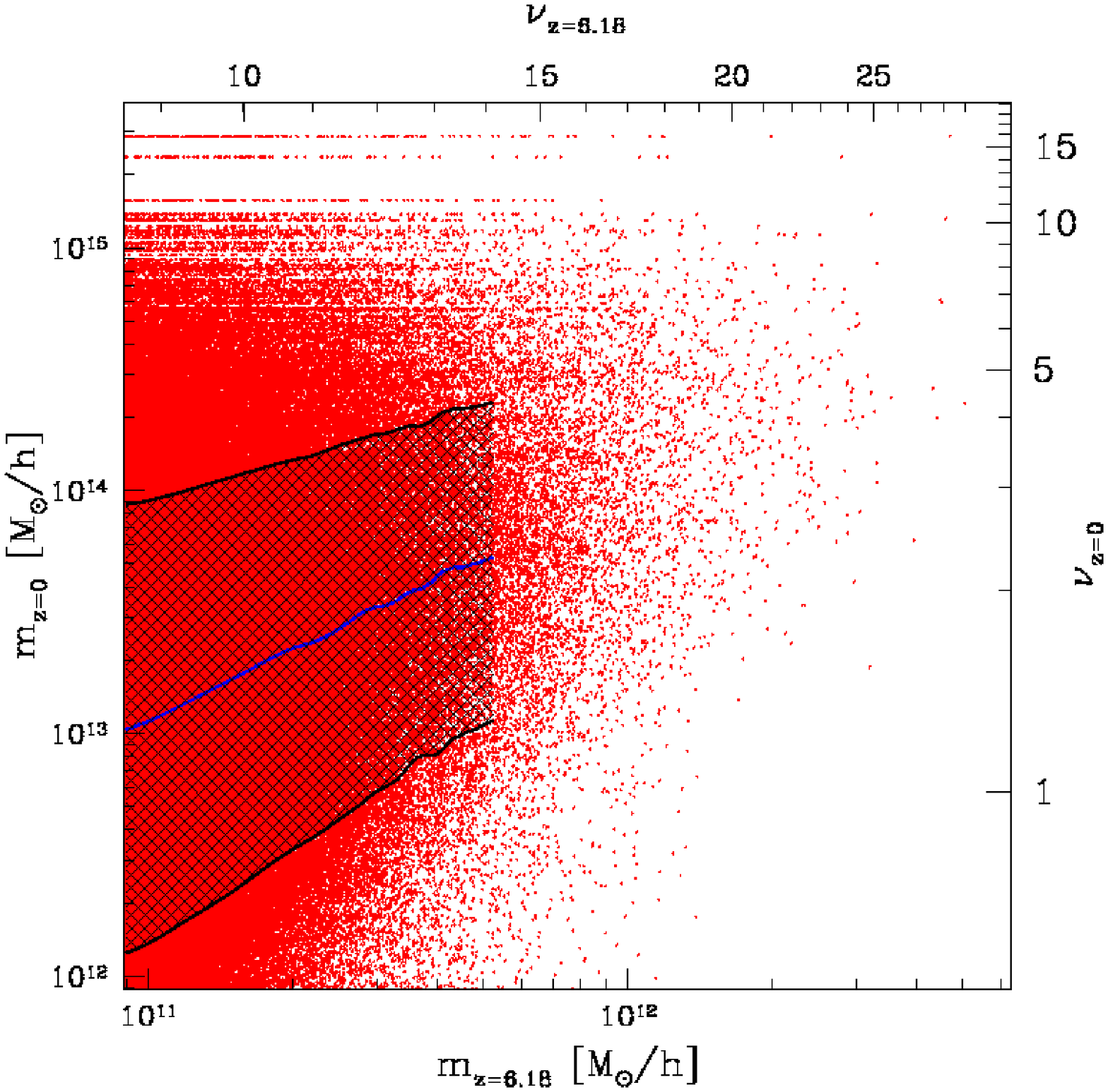} \caption{Mass at $z=0$ of the descendants
  of the most massive dark matter halos at $z=6.18$ in the Millennium
  Run \citep{MR05}. The shaded area within the region with the highest
  density of points represents for a given $z=6.18$ mass the 68\%
  confidenence level interval for the mass of the $z=0$ descendant,
  while the blue central line is the median of the distribution. Some
  of the most massive halos at $z>6$ have accreted only relatively
  little mass and a considerable scatter is present. The upper and
  right axes in the plot translate the halo mass into the
  dimensionless variable $\nu = \delta_c^2/ \sigma^2(M)$, used in the extended
  Press-Schechter formalism.}\label{fig:MR}
\end{figure}

From the Millennium Catalog\footnote{http://www.mpa-garching.mpg.de/millennium} we select all dark matter halos with more
than $10^3$ particles (corresponding to a minimum halo mass of $8.6
\cdot 10^{11}~h^{-1} \mathrm{M_{\sun}}$) in the $z=0$ snapshot and all
halos with more than $100$ particles in the $z=6.18$ snapshot. Within
these catalogs we also identify the descendants at $z=0$ of the
$z=6.18$ halos and plot them in Fig.~\ref{fig:MR}. The figure
immediately shows that the descendants of the most massive halos at
$z>6$ live in a variety of environments at $z=0$. Some of the most
massive halos at $z=6.18$ have accreted relatively little mass by
$z=0$ (a factor of a few of their initial mass), while others have
increased their masses by more than hundred times, reaching
$M>10^{15}~h^{-1} \mathrm{M_{\sun}}$. The scatter is substantial. In
particular the descendant of the most massive $z=6.18$ halo in this
simulation has a mass of only $2.2 \cdot 10^{14}~h^{-1}
\mathrm{M_{\sun}}$ at $z=0$, to be compared with the largest cluster
in the box, which has $M=3 \cdot 10^{15}~h^{-1}
\mathrm{M_{\sun}}$. Fig.~\ref{fig:MR} also quantifies the relation
between progenitor and descendant halos in terms of the dimensionless
variable $\nu = \delta_{c}^2 / \sigma^2(M)$ used in extended Press-Schechter
modeling. Here $\sigma^2(M)$ is the variance of density
fluctuations over a scale that contains a mass $M$ and $\delta_c(z)$
is the critical value of an overdensity in linear theory at redshift
$z$.

\begin{figure} 
  \plotone{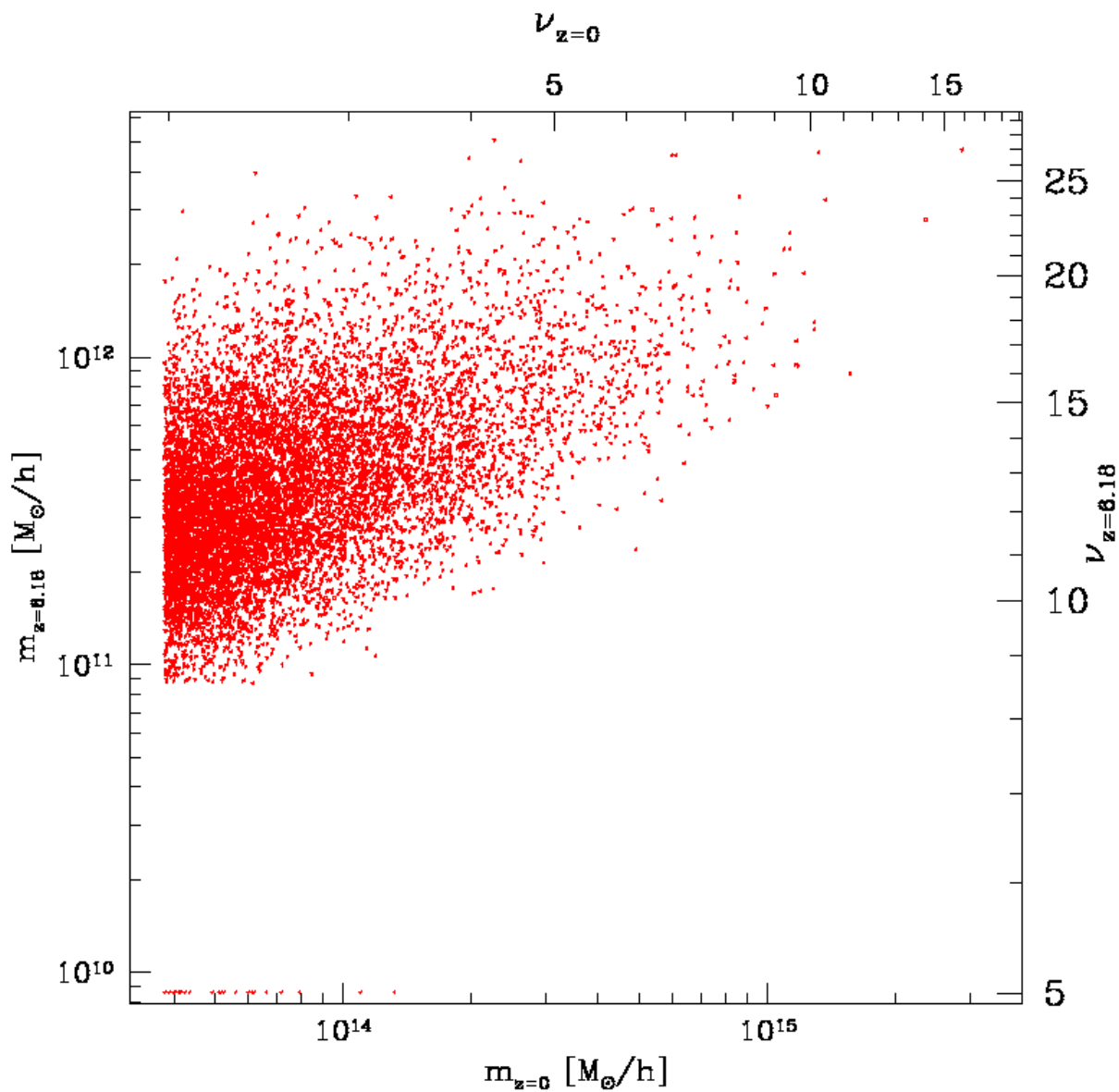} \caption{Mass at redshift $z=6.18$ of the most
  massive progenitor of the most massive halos at $z=0$ for the
  Millennium Run \citep{MR05}. $\nu$ values are given as in
  Fig.~\ref{fig:MR}. The few points with $m_{z=6.18} = 8.5 \cdot 10^9
  h^{-1} \mathrm{M_{\sun}}$ are associated to those $z=0$ halos that
  do not have a $z=6.18$ progenitor identified in the simulation (that
  is its below the minimum halo mass of $8.5 \cdot 10^{10} h^{-1}
  \mathrm{M_{\sun}}$ --- corresponding to $100$ particles). Given the
  considerable scatter in these plots, adaptive refinement of the most
  massive $z=0$ halo in a simulation would not necessarily lead to the
  most massive $z \approx 6$ halo of that box.  }\label{fig:MR1}
\end{figure}

\begin{figure} 
\plotone{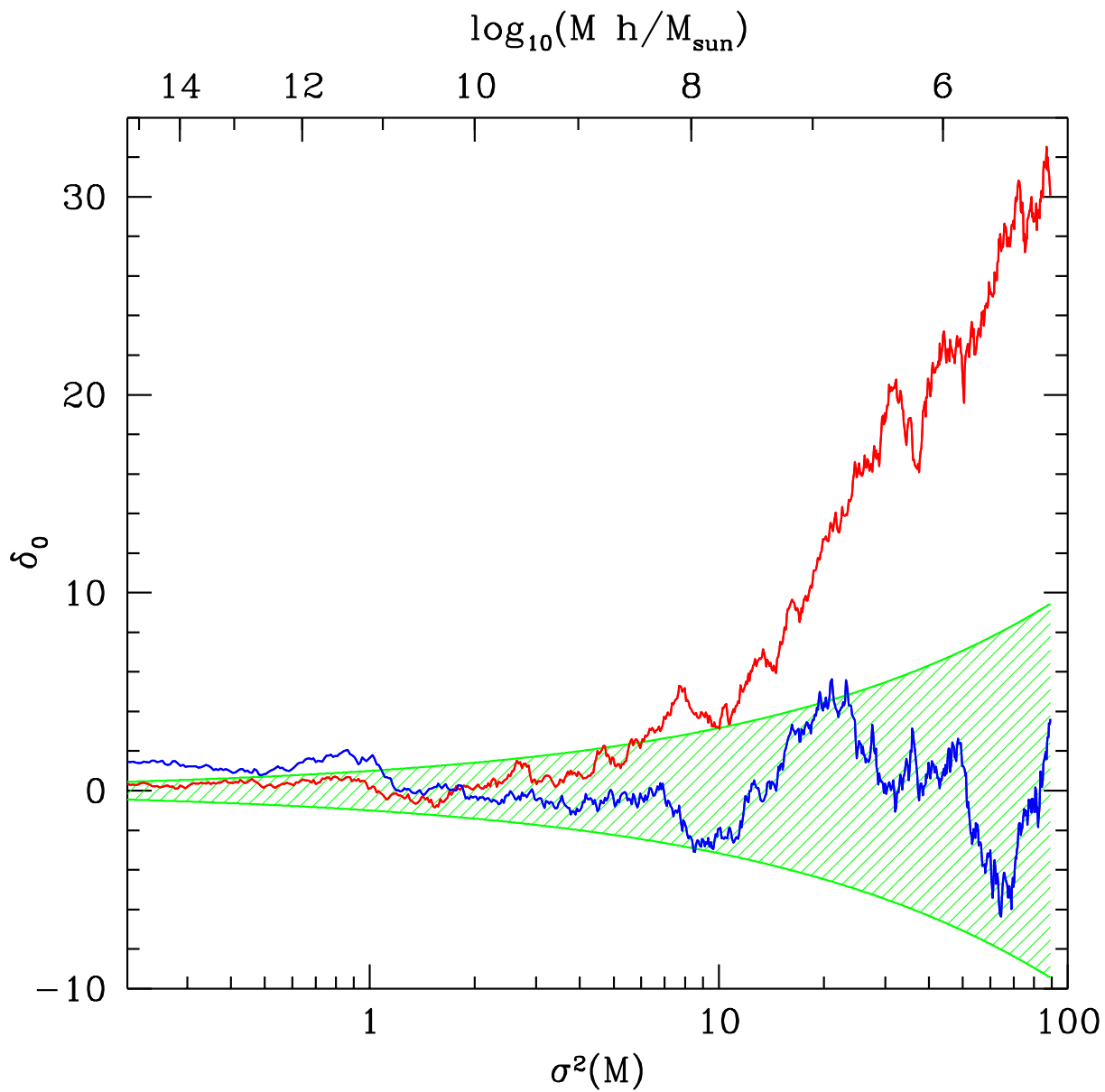}\caption{Random walks generated using a Millennium run
cosmology. The green shaded area represents for any value of
$\sigma^2(M)$ --- evaluated at $z=0$ --- the area enclosing 68\% of
the walks and thus has amplitude $\pm \sigma(M)$. The red and blue
walks have been selected from $10^3$ random realizations as (i) red:
the walk with the largest overdensity at $M=3 \cdot 10^5 h^{-1}
\mathrm{M_{\sun}}$ and (ii) blue: the walk with the largest
overdensity at $M=2 \cdot 10^{15} h^{-1} \mathrm{M_{\sun}}$.
}\label{fig:random_walk}
\end{figure}

A complementary picture is given by considering the mass distribution
of the most massive $z=6.18$ progenitor for the $10000$ largest $z=0$
halos, that is with a mass $M \gtrsim 4 \cdot 10^{13} h^{-1}
\mathrm{M_{\sun}}$ (Fig.~\ref{fig:MR1}). Again a considerable scatter
is present in the plot, with only a modest correlation between
descendant and progenitor mass. These results reflect the fact that
there are additional contributions to the density fluctuation power
spectrum over a scale $M_1$ compared to a scale $M_2>M_1$. This can be
also illustrated by random walks generated using a $\Lambda CDM$ power
spectrum (see Fig.~\ref{fig:random_walk}). The typical fluctuations at
small scales greatly exceed those at large scale, so that a walk with
excess power at small scales often has a rather typical amplitude 
at large scales, and vice versa.

\begin{figure} 
  \plotone{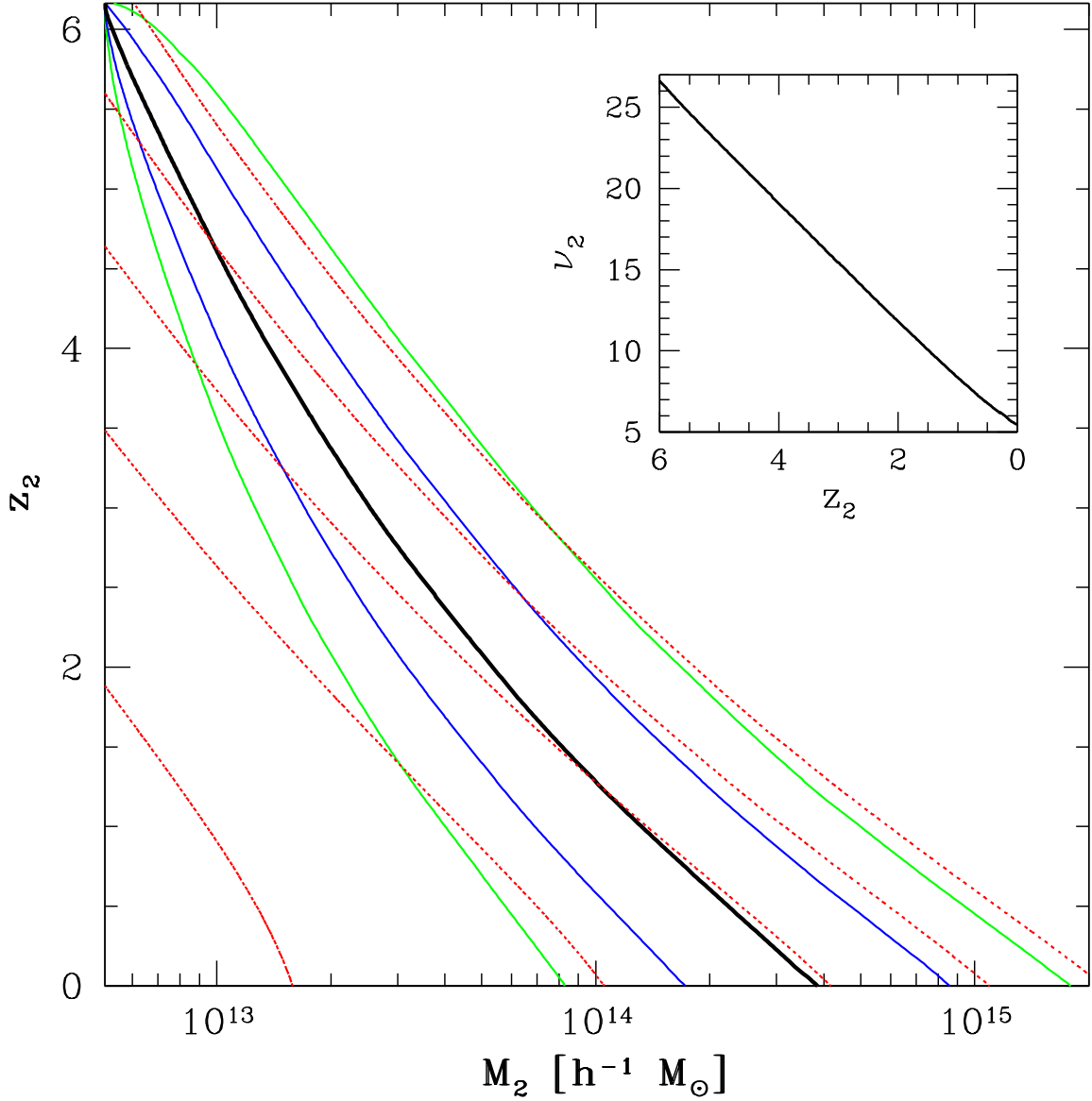} \caption{Main panel: Contours for the probability
  that a halo of mass $M_1=5 \cdot 10^{12} h^{-1} \mathrm{M_{\sun}}$
  at redshift $z_1=6.18$ is more massive than $M_2$ at redshift
  $z_2<z_1$ [$P(M>M_2,z_2|M_1,z_1)$]. The black line is the median of
  the distribution, the blue lines the $1 \sigma$ confidence levels
  and the green lines the $2 \sigma$ confidence levels. Red dotted
  lines: contour lines for the number density of dark matter halos in
  a $10^8 (\mathrm{Mpc}/h)^3$ volume (1,10,100,1000,10000 halos from
  top to bottom), obtained using the Sheth \& Tormen mass
  function. Small inset: Evolution of the main panel median line
  represented in terms of the extended Press-Schechter variable
  $\nu$. From this inset it is immediately clear that the
  descendant of a rare density peak progressively becomes more
  common.}\label{fig:EPS}
\end{figure}

A quantitative model for this behavior is available in the context of
extended Press-Schechter theory, for example by Eqs. 2.15 and 2.16 in
\citet{lacey93}. Using this formalism one can compute the probability
that a dark matter halo of mass $M_1$ at redshift $z_1$ evolves into a
halo of mass $M>M_2$ at $z_2<z_1$ [$P(M>M_2,z_2|M_1,z_1)$]. This is
shown in Fig.~\ref{fig:EPS}, where we plot the contour lines of $P$
for a $M_1=5 \cdot 10^{12}~h^{-1} \mathrm{M_{\sun}}$ halo at
$z_1=6.18$. The median of the distribution at $z=0$ is $M=4 \cdot
10^{14}~h^{-1} \mathrm{M_{\sun}}$ and at the 68\% confidence level
interval is $M \in [1.8:8.8] \cdot 10^{14}~h^{-1}
\mathrm{M_{\sun}}$. In this respect the Millennium Run is typical as
it lies within the $1 \sigma$ interval. From Fig.~\ref{fig:EPS} one
can also immediately see that it is relatively improbable for the
descendant of the most massive halo at $z=6.18$ to be the most massive
halo at $z=0$. In fact there is only a probability $p<2\%$ that the
mass of the descendant halo is above $2 \cdot 10^{15}~h^{-1}
\mathrm{M_{\sun}}$. The Millennium Run has a volume large enough to
contain $z=0$ halos more massive than $2 \cdot 10^{15}~h^{-1}
\mathrm{M_{\sun}}$ (in fact there are two of them), thus it is
expected that in less than $2\%$ of Millennium-like realizations the
most massive $z \approx 6$ halo is the progenitor of the most massive
$z=0$ halo. As a reference we provide the number density contours from
the \citet{she99} mass function, integrated to obtain the number of
objects above that mass at that redshift in the volume of the
Millennium Run (red dotted lines in Fig.~\ref{fig:EPS}): when $z
\lesssim 2.5$, the upper 2$\sigma$ confidence level contour lies at a
lower mass than the contour for the number density ($n=1$) associated
with the most massive dark matter halo in the volume. This mean that
the typical mass of the descendants progressively shifts toward that
of more common halos as the redshift decreases, as it is also
immediately visible when looking at the redshift evolution of the
value of $\nu$ for the median descendant, which decreases
progressively (small panel of fig.~\ref{fig:EPS}).

\section{The very first PopIII stars: where are they today?} \label{sec:PopIII}

A similar scenario holds for the relation between the location of the
very first PopIII stars, formed in dark matter halos with $M \approx
10^6 \mathrm{M_{\sun}}$ at $z>40$, and that of dark matter halos at
$z=0$. Here there is even less correlation than in the case of $z=6$
QSO halos (see Sec.~\ref{sec:QSOs}) because the halo mass dynamic
range involved is larger and thus there is an additional contribution
to the scatter from modes at small scales in the density fluctuation
power spectrum. Using a simulation volume of $\approx 1~\mathrm{Gpc}^3$,
\citet{ts07a} showed that the remnants of the very first PopIII stars
formed in dark matter halos of mass $10^6~h^{-1} \mathrm{M_{\sun}}$ at
$z>40$ end up at $z=0$ in dark matter halos with a median mass of $3
\cdot 10^{13}~h^{-1} \mathrm{M_{\sun}}$, about two order of magnitude
smaller than the largest halo in the simulation box. Here we confirm
and extend their result using extended Press-Schechter modeling.

\begin{figure} 
  \plotone{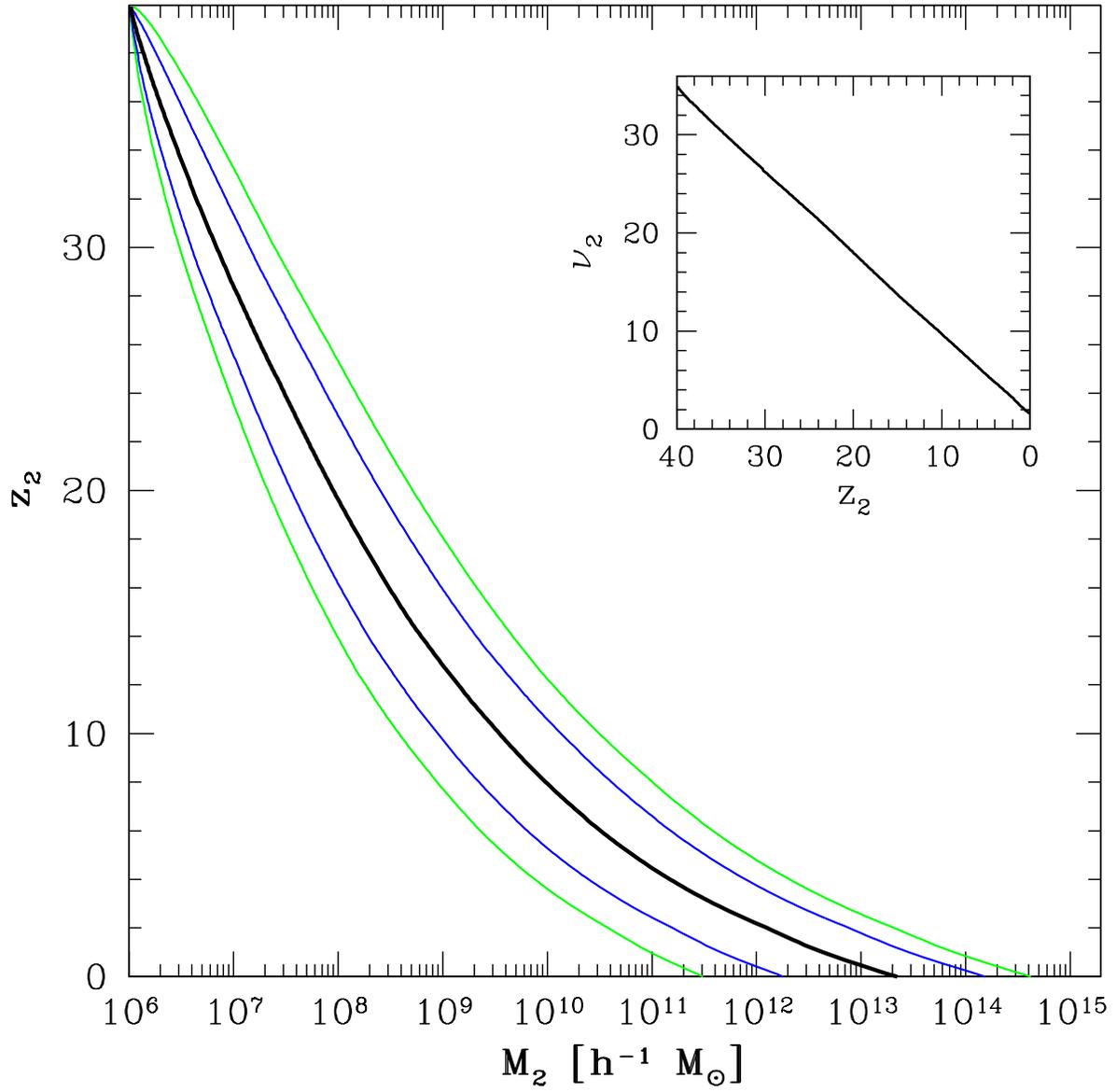} \caption{As in fig.~\ref{fig:EPS} but for a
  dark matter halo of mass $M_1= 10^{6} h^{-1} \mathrm{M_{\sun}} $
  formed at $z_1=40$.}\label{fig:EPS_PopIII}
\end{figure}

Using Eq. 2.15 and 2.16 of \citet{lacey93}, we compute at different
redshifts the probability distribution for the mass of the descendant
of a $10^6~h^{-1} \mathrm{M_{\sun}}$ dark matter halo formed at
$z_1=40$. The results are shown in Fig.~\ref{fig:EPS_PopIII}, where we
plot the median descendant mass versus the redshift $z_2$ (black
line), the $1 \sigma$ confidence level contours (blue lines) and $2
\sigma$ confidence level contours (green lines). These analytic
results agree well with the numerical simulations in \citet{ts07a} and
confirm that at $z=0$ the typical remnant of one of the very first
PopIII halos does not live in a supercluster, but is rather hiding in
a more common group environment. This is, e.g., contrary to the
conclusions ( ``The very oldest stars should be found today in the
central regions of rich galaxy clusters'') that \citet{whi00} drew
from the resimulation of a massive galaxy cluster. Similarly, at $z=6$
the remnant of a typical $z \geq 40$ PopIII star does not live in the
largest dark matter halos of that time, but rather has seeded a dark
matter halo of mass $3-4 \cdot 10^{10}~h^{-1} \mathrm{M_{\sun}}$,
typical for the faint $z\approx 6$ galaxies observed in deep surveys
such as the Hubble Space Telescope Ultra Deep Field \citep{ts08}.

%%%%%%%% Discussion about more common PopIII stars
These results have been obtained for very rare PopIII stars formed at
$z>40$. Extended Press-Schechter modeling predicts that the remnants
of more common PopIII stars formed at lower redshift live in
yet lower mass dark matter halos. For example a $10^6~h^{-1} \mathrm{M_{\sun}}$ dark matter halo
formed at $z=20$ has a $z=0$ descendant with median mass $4 \cdot
10^{12}~h^{-1} \mathrm{M_{\sun}}$. More massive, rarer, PopIII halos, with virial
temperature above $10^4$ K, can cool by neutral hydrogen rather than
molecular hydrogen so that their formation is less sensitive to
radiative feedback from other stars \citep[e.g. see][]{bromm04}. These
halos have a typical mass of $\approx 10^{8}~h^{-1} \mathrm{M_{\sun}}$ and if
they form at $z=20$, then their typical descendants are similar to
those of a $z=40$ $10^6 h^{-1} \mathrm{M_{\sun}}$ dark halo (see
Fig.~\ref{fig:EPS_PopIII}).

\section{Generation of Initial Conditions around Rare High-Redshift Halos}\label{sec:Ic}

Given the large scatter in the assembly histories of dark matter halos
is there an optimal way to select a region of a simulation box
centered around a rare high-redshift halo under the constraint of
limited computational resources?

\subsection{Analysis of the recursive resimulation method}

The method introduced by \citet{gao05} certainly presents a very
competitive advantage over a random selection of an equal subvolume
though it identifies a halo with an atypical
accretion history, biased toward having an above average merging rate
and living in an environment that tends to have an overdensity of
nearby halos \citep{bar04}. 
%This fact must be properly taken into account if the
%properties of the resimulation are then used to infer average
%properties of the galaxy population, especially if baryon physics
%and/or radiative feedback is added to the resimulation.
To better quantify the properties of halos identified by recursive
resimulation, we use the Monte Carlo method presented in
\citet{ts07a}, which is based on the identification of virialized dark
matter halos as density peaks in the linear density field. This
approach predicts well and without introducing systematic biases the
location and virialization redshift of the very first dark matter
halos when compared to the full non-linear dynamics of the simulation,
even though there are some statistical fluctuations on a halo-to-halo
basis \citep{bond96a,mes07}. For the comparison with \citet{gao05} we
adopt the following parameters from their paper: (i) $\Omega_M=0.3$,
$\Omega_{\lambda}=0.7$, $\Omega_b = 0.04$, $h=0.7$, $\sigma_8=0.9$,
spectral index $n_s=1$; (ii) parent box size edge $l_{box}=479$ Mpc
$h^{-1}$; (iii) largest dark matter halo in the box at $z=0$ of mass
$M_1=8.1 \cdot 10^{14} h^{-1} \mathrm{M_{\sun}}$; (iv) final
high-resolution simulation sphere of $1.25 h^{-1}$~Mpc (with volume
$V_5 = 8.18 h^{-3}$~Mpc$^3$); (v) the most massive halo in the final
resimulation region R5 is $M_{halo_{R5}} = 1.2 \cdot 10^5 h^{-1}
\mathrm{M_{\sun}}$ at $z=48.84$. Based on these assumptions, our
analysis yields the following results:
\begin{itemize}

\item The number density of dark matter halos of mass $M \geq
M_{halo_{R5}}$ is $n(M>M_{halo_{R5}}) = 0.17 h^3 $Mpc$^{-3}$
(Sheth-Tormen mass function) or $n(M>M_{halo_{R5}}) = 0.0025 h^3
$Mpc$^{-3}$ (Press-Schechter mass function) at $z=48.84$, so these
halos are not particularly rare. Specifically, in a random region of
volume $V_5$ at $z=48.84$, the expectation value for the number of
halos more massive than $M_{halo_{R5}}$ is greater than unity ($n(M>
M_{halo_{R5}}) \cdot V_5 = 1.39$) when using the \citet{she99} formula.
However, due to the high bias of the halos, $b \approx 16$, the actual
fraction of empty volumes as given by our Monte Carlo code is
$98.7\%$. Thus the \citet{gao05} method is a substantial improvement
over random selection of a volume $V_5$ within the parent box. 

\item The first halo of mass $M_{halo_{R5}} = 1.2 \cdot 10^5 h^{-1}
\mathrm{ M_{\sun}}$ in the \emph{whole} box virializes at $z>62$ at
99\% of confidence level. The possibility that the most massive halo
identified by \citet{gao05} in their final resimulated volume is the
most massive of the whole box at that redshift is ruled out at a
confidence level greater than $1-10^{-7}$ (confidence level limited by
the precision of the numerical integration).

\item The first halo of mass $M_{halo_{R5}}$ that ends up at $z=0$ in
a dark matter halo of mass $M_1=8.1 \cdot 10^{14} h^{-1} \mathrm{
M_{\sun}}$ is formed at $z>50.8$ at 99.9\% of confidence level. Thus
it is unlikely that the \citet{gao05} halo of mass $M_{halo_{R5}}$ at
$z=48.84$ is the most massive progenitor of the cluster that will
contain it at $z=0$.

\end{itemize} 

From this analysis it is indeed confirmed that recursive refinement
works well to select a sub-region for resimulation that host a rare
high-z dark matter halo, but still this halo is not one of the very
first of its kind in the box. If the interest is primarily in the
local physics around one of these rare structures, a viable
alternative is the constrained realization method
\citep{hf91,bert01}. A constrained realization does not require a
hierarchy of resimulations, but rather introduces the overdensity in
the initial conditions by construction, so that the realized initial
conditions do not carry information about its rarity or its typical
surrounding environment.

\subsection{Unbiased selection of rare halos: the density field method}

If one is instead interested in selecting one of the very first dark
matter halos in a given simulation volume, with a particular interest
in representative halos, that is halos with unbiased accretion
histories, we propose instead to select the final region of interest
from a density field at uniform resolution. When the dynamic range of
the resimulation is not too large, it is of course possible to
identify the region to be refined directly from the low-resolution
dark matter halo catalogs. If this is not possible, then the
resimulation volume can be identified from a high resolution, but
uniform, refinement of the density field. The idea is the following:
%%%%%%%%%%%%%%%
\begin{enumerate}

\item A high-resolution density field is generated over the whole box.
The mass of a field cell is that of the high-$z$ dark
matter halo of interest (e.g. $10^6 h^{-1}  \mathrm{ M_{\sun}} $ for a PopIII
halo). A technical, but important, detail is that a top-hat
filter in frequency space must be used when generating this field, as this
guarantees that the mass function obtained from peak analysis matches
that of \citet{PS}.

\item The highest peak in the high-resolution field is
identified. This is the location where one of the first halos at the
mass scale of the density field is formed within the \emph{whole} simulation
box.

\item A region around that peak can now be selected for resimulation,
with additional fluctuations added over the high resolution density
field, for example by using the GRAFIC2 refinement method of
\citet{bert01}. Outside the selected region the high resolution field
can be degraded if necessary to generate more massive particles.
\end{enumerate}
%%%%%%%%%%%%%%

This method does not necessarily require overwhelming computational
resources. For example, to select the largest $z=6$ dark matter halo
in a volume relevant for the study of bright high-$z$ QSOs, a mass
resolution of $\approx 5 \times 10^{11} h^{-1} \mathrm{ M_{\sun}} $ is
more than sufficient, which translates into a $N = 512^3$ density
field grid for a cosmic volume of $\approx (1 \mathrm{Gpc}/h)^3$. Such
a grid only requires 512MB of RAM for storing. Even identifying the
positions of the first PopIII stars in a large volume is not
unrealistic. A $N=2048^3$ density grid requires only 32GB of RAM and
provides a mass resolution of $1.2 \times 10^5 h^{-1}
\mathrm{ M_{\sun}} \equiv M_{halo_{R5}}$ (the first halo in
\citealt{gao05}) over a volume of $(23.74 h^{-1}
\mathrm{Mpc})^3$. Given the number density of such halos---$n \approx
0.11 h^3 \mathrm{Mpc}^{-3}$---about 1471 will have formed on average
in the box by $z=48.84$. In fact, using the Monte Carlo code of
\citet{ts07a} we predict that there is a probability $p< 10^{-3}$ of not
forming one of these halos before $z=48.84$. The median redshift of
formation for the first halo of mass $M_{halo_{R5}}$ is $z=51.6$. With
larger resources one can generate a $4096^3$ density grid over a
volume of $(100 \mathrm{Mpc}/h)^3$, obtaining a mass resolution of $10^6 h^{-1}
\mathrm{ M_{\sun}} $. This requires about 256GB of RAM, easily
available on a parallel computing cluster of moderate size. With a
top-end supercomputer one can use more than $10^{12}$ cells, with a
dynamic range that permits identification of smaller mass halos in the
same volume or an increase the volume of the parent box for the same
assumed PopIII halo mass.
 
Our proposed density field method has also another advantage: As we
generate a uniform, very high-resolution density grid over the whole
box, aliasing errors in the zoom refinement procedure are less severe
than if one follows the \citet{gao05} recursive refinement method (see
\citealt{bert01} for a detailed analysis of the errors introduced when
large zooming factors are used).

\subsection{Preliminary testing of the density field method}\label{sec:valid}

Detailed analysis and applications of our method to select
high-redshift halos is deferred to a follow-up paper. Here we present
a preliminary testing to assess its performance. For this we consider
a $N=512^3$ cosmological simulation in a box of edge $l_{box}=512
h^{-1} \mathrm{Mpc}$ (details on the simulation are discussed in
\citealt{ts07a}). We construct the halo catalog for a snapshot at
$z=5$ using the HOP halo finder \citep{eis98}, finding that the most
massive halo has a mass of $1.25 \cdot 10^{13} h^{-1}
\mathrm{M_{\sun}}$ ($187$ particles). The density field used to
generate the initial conditions for this $N=512^3$ run is then (i)
downgraded to a $N_{down}=128^3$ grid (where one low resolution cell
corresponds to 64 of the original grid cells) using the \citet{hf91}
constrained realization method and (ii) evolved in linear theory to
$z=5$. The highest density peaks in the low-resolution field are then
identified and their location compared with that of the dark matter
halos identified from the N-body simulation at full resolution. Within
the ten highest peaks (with linear overdensities from $\delta =1.84$
to $\delta = 1.75$), six of them are associated with dark matter halos
with more than $64$ particles, including all the top three
overdensities. The top overdensity is matched to the fourth most
massive halo of the snapshot, with $141$ particles (versus the $187$ of
the most massive). Of the remaining unmatched peaks, two are
associated with halos with less than 64 particles and two appear to be
still in the process of virialization, so the central density of their
groups in the N-body run does not qualify them as halos. The most
massive halo in the box is missed by the density field analysis
because its particles are spread over several adjacent cells in the
downgraded density field. This is an intrinsic limitation of our
method rooted in the use of a fixed grid in the position space, which
is bound to miss some of the the dark matter halos that are off-center
with respect to the spatial location of the grid cells. However this
only leads to miss a random fraction of the rare halos, without an
environmental bias as in the \citet{gao05} method. 

\section{Conclusion and Discussion}\label{sec:conc}

By means of both analysis of numerical simulations and of extended
Press-Schechter modeling we investigated the relation between the most
massive dark matter halos at different redshifts. The main conclusion
of this work is that---contrary to expectations from many recent works
(e.g. see \citealt{MR05,ree05,gao05,gao07,li07})---the most massive
halo at a redshift $z_1>z_2$ does not necessarily evolve into the most
massive at $z=z_2$. This is a robust conclusion that can be naturally
understood in the context of growth of dark matter density
perturbations, for example by constructing merger trees through the
\citet{lacey93} model. Rare high-redshift objects, such as the
remnants of the first PopIII stars and QSOs, are not hosted at $z=0$
in the most massive halos, but rather live in a variety of
environments. For example, the typical $z>40$ PopIII star remnant
lives in a dark matter halo that at $z=0$ has a mass of $\approx 2
\cdot 10^{13} h^{-1} \mathrm{M_{\sun}}$, typical for a galaxy group,
and not within rich clusters as claimed by \citet{whi00}. Similarly
the descendant of a typical $z\approx 6$ QSO is not located within the
most massive clusters at $z=0$ as assumed by \citet{li07}.

These conclusions have important consequences on the application of
the recursive simulation method introduced by \citet{gao05} to
identify high-redshift rare halos based on progressive refinement of
regions centered around the most massive $z=0$ cluster. In fact, while
the recursive method is indeed effective at identifying a sub-region
of the simulation with earlier-than-average structure formation, it
finds neither the earliest structures in the box, which are dominated
by small-scale density fluctuations, nor typical early structure, as
it preferentially identifies objects located in the regions with the
highest bias.
 
These limitations may have only a minor effect when the goal is to
investigate the formation of one rare Population III halo in the
simulation box, as it is done for example in \citet{gao05} and in
\citet{ree05}. However in different physical scenarios it is important
to correctly estimate the rarity of the halo simulated and to assess
how typical their growth histories are. This is critical if additional
physics beyond gravitational interactions is included, such as star
formation and radiative feedback. One such example is the formation of
rare high-redshift QSOs: \citet{li07} use the \citet{gao05} method to
identify at $z\approx 6.5$ ``the most massive halo in a $\approx$ 3
Gpc$^3$ volume'' and then conclude that the QSO formed in this halo
reproduces the properties of observed QSOs with the same number
density. From our analysis in Sec.~\ref{sec:QSOs} it is clear that the
halo identified by \citet{li07} as progenitor of the largest $z=0$
cluster is not likely to be the most massive at $z \approx 6.5$. Thus
other similar or more massive halos are expected to be present at
$z \approx 6.5$ in their $\approx$ 3 Gpc$^3$ simulation volume: in principle
any of these halos could host a bright QSOs, with important
consequences for the comparison between observed and simulated QSO
number densities. In addition, when the goal is to study the
environment in which QSOs live, selecting host halos with the
resimulation method introduces systematic effects in the results
difficult to quantify and correct for, because these halos would have
above-average growth (and merging) histories.

To avoid selecting only the halos with atypical accretion histories
and in an attempt to improve over the identification of some of the
rarest high-redshift halos in a box, we propose instead to select the
initial conditions for high-resolution resimulation based on the
analysis of the linear density field at uniform resolution. The
method, described in Sec.~\ref{sec:Ic}, identifies subregions of the
simulation box with high-redshift halos as those with the highest
peaks in the density field, requiring a mass resolution in the field
comparable to that of the mass of the halos one wishes to select.  The
applicability of the method is thus limited only by the size of the
largest density grid that can be accommodated in the available memory,
otherwise requiring only a modest amount of computing time compared to the
\citet{gao05} method. This is because our method bypasses the need of
a series of N-body simulations to be carried out in addition to the
final run.

Unfortunately, our method does not guarantee identification of \emph{the
first} halo on the desired mass scale, as highlighted by some preliminary
testing we presented in Sec.~\ref{sec:valid}. This seems still an
elusive goal. When the density field is defined over a fixed grid,
there is not a perfect match between the halo catalog constructed from
the density field and that obtained by increasing the resolution of
the field and then following the full non-linear dynamics with an
N-body simulation. An extensive validation of our linear density field
initial conditions generation and its application to the formation of the first
bright QSOs will be discussed in a subsequent paper.

\acknowledgements

This work was supported in part by NASA grants JWST IDS NAG5-12458 and
HST-GO10632. We thank the referee for useful comments and
suggestions. We are grateful to Gerard Lemson and to the Millennium
Run collaboration for allowing us to use their merger tree catalogs.
 
%%%%%%%%%%%%%%%%%%%%%%%%%%%%%%%%% 

\clearpage

%%%%%%%%%%%%%%%%%%%%%%%%%%%%%%%%%%%%%%%%%%%%

%%%%%%%%%%%%%%%%%%%%%%%%%%%%%%%%%%%%%%%%%%%%

\clearpage

%%%%%%%%%%%%%%%%%%%%%%%%%%%%%%%%%%%%%%%%%%%%

%%%%%%%%%%%%%%%%%%%%%%%%%%%%%%%%%%%%%%%%%%%%

%%%%%%%%%%%%%%%%%%%%%%%%%%%%%%%%%%%%%%%%%%%%

%%%%%%%%%%%%%%%%%%%%%%%%%%%%%%%%%%%%%%%%%%%

%%%%%%%%%%%%%%%%%%%%%%%%%%%%%%%%%%%%%%%%%%%%

%%%%%%%%%%%%%%%%%%%%%%%%%%%%%%%%%%%%%%%%%%%

\end{document}